\documentclass{article}
\usepackage{spconf,amsmath,graphicx}
\usepackage{multirow,amsfonts,cite,array,url}
\usepackage{color}
\usepackage{booktabs}

\title{AIMDiT: Modality Augmentation and Interaction via Multimodal Dimension Transformation for Emotion Recognition in Conversations}
\name{
    Sheng Wu$^{1,2}$  \quad
    Jiaxing Liu$^2$ \quad 
    Longbiao Wang$^{2,*}$ \thanks{* Corresponding author} \quad 
    Dongxiao He$^2$ \quad 
    Xiaobao Wang$^2$ \quad 
    Jianwu Dang$^2$}
\address{
    $^1$ School of New Media and Communication, Tianjin University, Tianjin, China  \\
    $^2$ Tianjin Key Laboratory of Cognitive Computing and Application, \\
    College of Intelligence and Computing, Tianjin University, Tianjin, China \\
}
\begin{document}
%
\maketitle
\begin{abstract}
Emotion Recognition in Conversations (ERC) is a popular task in natural language processing, which aims to recognize the emotional state of the speaker in conversations. 
While current research primarily emphasizes contextual modeling, there exists a dearth of investigation into effective multimodal fusion methods. 
We propose a novel framework called AIMDiT to solve the problem of multimodal fusion of deep features.
Specifically, we design a Modality Augmentation Network which performs rich representation learning through dimension transformation of different modalities and parameter-efficient inception block. On the other hand, the Modality Interaction Network performs interaction fusion of extracted inter-modal features and intra-modal features. 
Experiments conducted using our AIMDiT framework on the public benchmark dataset MELD reveal 2.34\% and 2.87\% improvements in terms of the Acc-7 and w-F1 metrics compared to the state-of-the-art (SOTA) models.
\end{abstract}

\section{Introduction}
\label{sec:intro}

Emotion recognition in conversations (ERC) is a novel task within the realm of natural language processing, holding a pivotal role in human-computer interaction \cite{DBLP:conf/CRC/BraveN07} and social media analysis \cite{DBLP:conf/IWSE/ChatterjeeNJA19}. 
There is an increasing focus on multimodal ERC, which involves leveraging text, audio, and visual information to achieve more accurate emotion recognition.


Multimodal data tends to be asynchronous, with distributional gaps in text, audio, and visual modal data, which tends to introduce information redundancy \cite{DBLP:conf/mm/YangKHZ22, DBLP:conf/cvpr/LvCHDL21}. In ERC, most of the work focuses on context modeling based on sequence models or graph models and often ignores the work on multimodal fusion 
\cite{DBLP:conf/acl/HuWH20, DBLP:conf/aaai/MajumderPHMGC19, DBLP:conf/emnlp/GhosalMPCG19}. Although these early works can realize multimodal data fusion by simple early-late fusion \cite{DBLP:conf/emnlp/ZadehCPCM17, DBLP:conf/acl/MorencyLZLSL18, DBLP:conf/icmcs/FuOWGSLD21, DBLP:conf/acl/LiTZZ22}, they cannot extract the inter-modal interaction information and the simple concatenation operation makes it easy to introduce redundant information. In recent years, some ERC works based on multimodal fusion of depth graphs have been proposed \cite{DBLP:conf/acl/HuLZJ21}, but due to the characteristics of depth graphs, redundant information gradually accumulates in the vector space of each layer and is prone to over-smoothing problem that makes node differentiation insufficient \cite{DBLP:conf/icassp/HuHWJM22}.

We have observed subtle distinctions in the distribution of text, audio, and visual information in emotional expressions, but they tend to be concentrated over a brief time span. For the purpose of discussion, we standardize the embedding dimensions of various modalities, treating vector length as the temporal dimension. Consequently, modal vectors can be conceptualized as 1D time series. Drawing inspiration from \cite{DBLP:conf/mm/HazarikaZP20, DBLP:conf/mm/YangHKDZ22, DBLP:conf/ICLR/WuHLZWL23}, we refer to changes in adjacent time intervals within a single modality and the same time interval across different modalities as intra-modal variation and inter-modal variation, respectively. To represent these characteristics, we transform 1D vectors into 2D, where the length axis signifies intra-modal information and the dimensional axis signifies inter-modal information. As a result, we successfully incorporate intra- and inter-modal features into 2D tensors.

\begin{figure*}[t]
  \centering
  \includegraphics[width=0.93\linewidth]{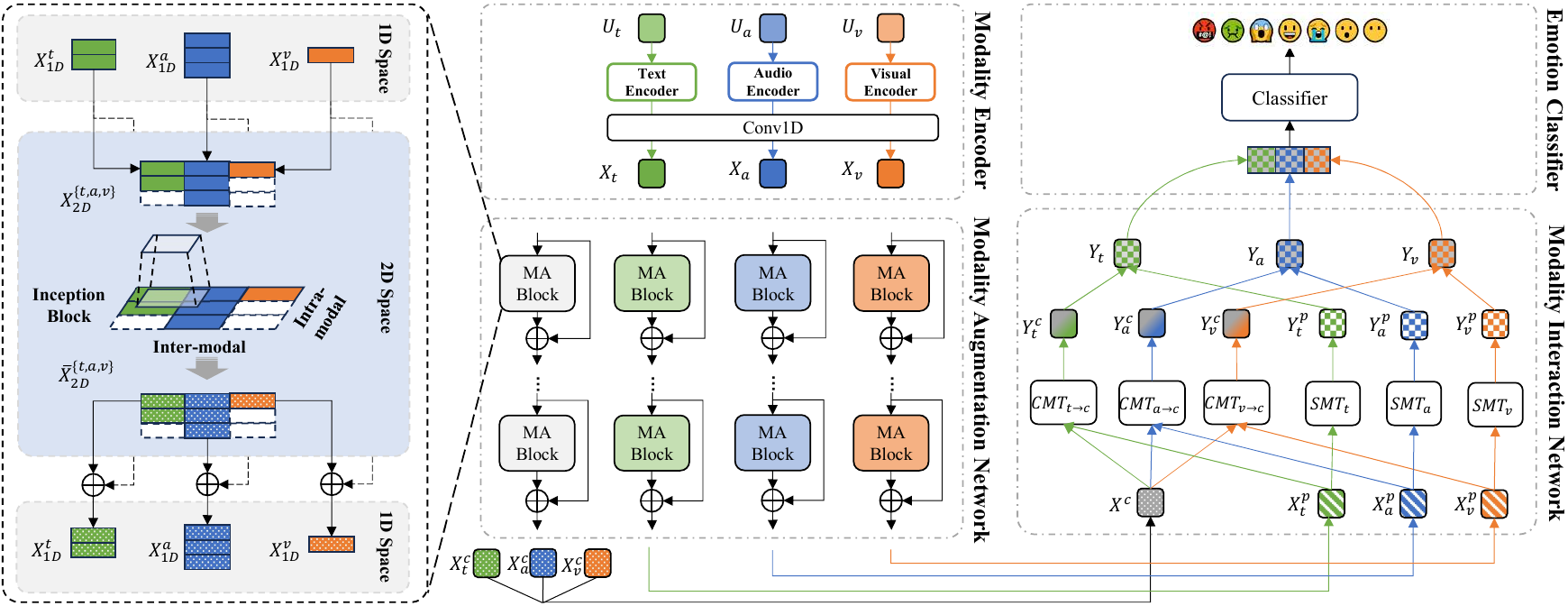}
  \caption{Framework illustration of the AIMDiT based emotion recognition in conversations, which consists of four key components: Modality Encoder, Modality Augmentation Network, Modality Interaction Network, Emotion Classifier.
 }
  \label{fig:overall}
\end{figure*}

Based on the above motivation, we propose the Modality Augmentation and Interaction Network via Dimension Transformation (AIMDiT). 
We first design the Modality Augmentation Network (MAN) based on dimension change and convolution by Inception. Specifically, the MAN consists of multiple layers of MABlock with residual links, which captures intra- and inter-modal information in 2D space. In order to perform an effective fusion of modalities, inspired by \cite{DBLP:conf/icassp/TsaiBLKMS19}, we propose the Interaction Network (MIN) where inter-modal feature and intra-modal feature guide the fusion of each other. Finally, emotion classifier predicts the emotion label.

The work of this paper can be summarized as follows: 1) We propose a multimodal fusion model named AIMDiT to facilitate understanding in ERC. 2) We design a network called MAN with dimension transformation and Inception convolution, which can effectively learn intra-modal and inter-modal features. 3) We introduce a fusion network called MIN that utilizes different types of modality features to fuse modalities. 4) Extensive experiments on benchmark datasets demonstrate the effectiveness and superiority of the proposed model.

\section{PROPOSED METHOD}
\label{sec:method}


The proposed model AIMDiT in this paper is shown in Fig.~\ref{fig:overall}. A primary utterance involves three modalities: $U_t$ (text), $U_a$ (audio), and $U_v$ (visual). Initially, we extract features from these modalities, yielding $X_t$,$X_a$,$X_v$. To capture inter- and intra-modal features, we devise the Modality Augmentation Network, elevating $X_t$, $X_a$, $X_v$ to higher dimensions for richer information. Following this, the Modality Interaction Network interacts with the MAN's output, guided by a specific direction. This leads to subsequent emotion classification. Further details are outlined below.

\subsection{Modality Augmentation Network (MAN)}

For the same modality, each time point involves both adjacent regions inside a modality and in-phase changes between different modalities, that is, changes within modalities and changes between modals. To capture both features at the same time, we build a module Modality Augmentation Block (MABlock), which explores rich information between different modalitys through dimensional changes.

Specifically, ignoring the dimension $d$, we consider $t$, $a$ and $v$,  where each modal corresponds to having 1D sequences $X^{t}_{1D} \in \mathbb{R}^{T_t \times d}, X^{a}_{1D} \in \mathbb{R}^{T_a \times d}, X^{v}_{1D} \in \mathbb{R}^{T_v \times d}$, and  with lengths $T_{t}, T_{a}, T_{v}$, respectively. We can then convert the three tensors into a 2D tensor by using the following formula:

\begin{equation}
\resizebox{0.9\linewidth}{!}{$
  \mathbf{X}^{t, a, v}_{2D} = Reshape(Padding({X^{t}_{1D},X^{a}_{1D},X^{v}_{1D}})) \in \mathbb{R}^{T_\delta \times 3d}, 
  $}
\end{equation} 
where $Padding(\cdot)$ is the process of padding each of the three tensors into a new tensor of uniform length, and use the length of the longest tensor of the three as the length of the new tensor $T_\delta = max(T_t, T_a, T_v)$.
Eventually we obtain a 2D tensor ${X}^{t,a,v}_{2D}$ .
After the transformation, we process the 2D tensor by inception block with multiple 2D kernels of different scales to obtain a 2D tensor $\overline{X}^{t,a,v}_{2D}$ with global information:
\begin{equation}
   \overline{\mathbf{X}}^{t,a,v}_{2D} = Inception({X}^{t, a, v}_{2D}).
\end{equation} 

Since the 2D kernel of the inception block is multiscale, it can capture information at adjacent time points within each modality and at the same time point for different modalities, and is able to aggregate intra- and inter-modal variations at different scales. 
Subsequently, we reconvert the learned $\overline{X}^{t,a,v}_{2D}$ tensor into 1D space, and truncate the three tensors with length ${T}_{\delta}$ back to their original lengths:

\begin{equation}
  \overline{\mathbf{X}}^{t}_{1D},\overline{\mathbf{X}}^{a}_{1D},\overline{\mathbf{X}}^{v}_{1D} = Debulk(Reshape(\overline{\mathbf{X}}^{t, a, v}_{2D})),
\end{equation} 
where $\overline{X}^{t}_{1D} \in \mathbb{R}^{T_\delta \times d}, \overline{X}^{a}_{1D} \in \mathbb{R}^{T_\delta \times d}, \overline{X}^{v}_{1D} \in \mathbb{R}^{T_\delta \times d}$.


As shown in Fig. 1, we connect the Modality Augmentation Block (MABlock) with residuals to form a MAN in order to prevent the problem of gradient vanishing in deep networks and to improve the expressive power of the network. Specifically, given three ${X}^{t}_{1D}, {X}^{a}_{1D}$, and ${X}^{v}_{1D}$ with lengths ${T}_{t},{T}_{a},{T}_{v}$, respectively, which are represented by a one-dimensional tensor ${X}_{1D}={\{X}^{t}_{1D}, {X}^{a}_{1D}, {X}^{v}_{1D}\}$ with length ${T}_{\delta}$, and for the $k$th-layer MAN, the inputs are ${X}^{k-1}_{1D}$, and its computation procedure is represented by Equation 4:
\begin{equation}
  \mathbf{X}^{k}_{1D} = MABlock(\mathbf{X}^{k-1}_{1D} ) + \mathbf{X}^{k-1}_{1D}. 
\end{equation} 

According to Eqs. 1, 2, and 3 the whole process of the $k$th MAB can be summarized as follows:
\begin{equation}
  \begin{split}
    \mathbf{X}^{k}_{2D} &= Reshape(Padding(\mathbf{X}^{k-1}_{1D})), \\  
    \overline{\mathbf{X}}^{k}_{2D} &= Inception(\mathbf{X}^{k}_{2D}), \\
    \mathbf{X}^{k}_{1D} = \overline{\mathbf{X}}^{k}_{1D} &= Debulk(Reshape(\overline{\mathbf{X}}^{k}_{2D})).
  \end{split}
\end{equation}

After the processing of the k-layer MAB, we get three modal outputs $\overline{X}^{t}_{1D},\overline{X}^{a}_{1D},\overline{X}^{v}_{1D}$, and finally we need to fuse these three outputs, where we use the average fusion. Through the above operation, we can extract to the inter-modal Feature $ \mathbf{{X}^{c}} = mean(\overline{X}^{t}_{1D},\overline{X}^{a}_{1D},\overline{X}^{v}_{1D})$. 

In order to capture the intra-modal feature ${X}^{p}_{\alpha}$ within the modality $\alpha$ as well as to maintain the consistency of the MAN structure, we take the text, audio and visual feature inputs as the inputs to the MAN respectively, i.e., three MANs are used to process each of the three modalities, which are denoted by ${X}^{p}_{\alpha} = MAN(\overline{\mathbf{X}}^{\alpha}_{1D}, \overline{\mathbf{X}}^{\alpha}_{1D}, \overline{\mathbf{X}}^{\alpha}_{1D}), {\alpha}={\{t,a,v\}}$.


	

\subsection{Modality Interaction Network (MIN)}
In order to fully utilize the consistent and complementary information in the acquired inter-modal and intra-modal features, we propose modality interaction network, which consists of cross-modal Transformer and self-modal Transformer \cite{DBLP:conf/nips/VaswaniSPUJGKP17}, with the intra-modal features guiding the reinforcement of the inter-modal feature.

We apply the cross-modal attention proposed by \cite{DBLP:conf/icassp/TsaiBLKMS19}. 
The main idea is to use modality $\beta$ to reconstruct modality $\alpha$, so as to achieve the fusion of the two modalities.
Specifically, the modal $\beta$ tensor ${X}_\beta$ passes through as the inputs of $K$ and $V$ in the attention, and the modal $\alpha$ as the input of $Q$. Then, the cross-modal attention ${Y}_{\alpha}$ is defined as follows:


\begin{equation}
\begin{aligned}
  {Y}_{\alpha} = softmax(\frac{{X}_{\alpha}{W}_{{Q}_{\alpha}}{W}^{T}_{{K}_{\beta}}{X}^{T}_{\beta}}{\sqrt{{d}_{k}}}){X}_{\beta}{W}_{{V}_{\beta}}.
\end{aligned}
\end{equation} 

Based on Cross-modal Attention, we constitute Cross-modal Transformer by stacking this structure. 
 CMT consists of k layers of Cross-modal Transformer, which takes the outputs ${X}^{i-1}_{{\beta}\rightarrow{\alpha}}$ and ${X}_{\beta}$ of the previous layer as inputs, and the outputs are the enhanced features ${X}^{i}_{{\beta}\rightarrow{\alpha}}$:
\begin{equation}
  {X}^{i}_{\beta\rightarrow\alpha}=\mathbf{CMT}^{i}_{\beta\rightarrow\alpha}({X}^{i-1}_{\beta\rightarrow\alpha},{X}_{\beta}),
\end{equation}
where ${\beta} \in {\{t,a,v\}}$, ${X}^{i}_{\beta\rightarrow\alpha} \in \mathbb{R}^{{T}_{\delta} \times d}$.
In addition, in order to fuse the inter-modal and intra-modal information, we also performed the traditional self-attention based Selfmodal Transformer processing for each modality, and finally averaged the two to obtain a comprehensive representation of each modality. 
It is expressed as follows:
\begin{equation} 
\resizebox{0.7\linewidth}{!}{$
  \begin{aligned}
     {Y}_{\alpha} &= \frac{{Y}^{c}_{\alpha} + {Y}^{p}_{\alpha}}{2} \\ 
      &= \frac{\mathbf{CMT}_{\alpha \rightarrow c}({X}^{c},{X}^{p}_{\alpha}) + \mathbf{SMT}_{\alpha}({X}^{p}_{\alpha}, {X}^{p}_{\alpha})}{2}, 
    \end{aligned}
    $}
\end{equation}
where ${Y}^{c}_{\alpha}$ is the output obtained by CMT using the inter-modal feature ${X}^{c}$ and the intra-modal feature of the modal $\alpha$. ${Y}^{p}_{\alpha}$ is the output obtained by SMT for the modal ${\alpha}$, where ${\alpha} \in {\{t,a,v\}}$.

\subsection{Emotion Classifier} \label{sec:emo}
The Emotion classifier uses connected multivariables as inputs to perform sentiment prediction.
Finally we input ${Y}_{tav}$ to the softmax layer to obtain the sentiment category:
\begin{equation}
  \begin{split}
  {P}_{tav} &= GELU({W}_{1}({Y}_{t} \oplus {Y}_{a} \oplus {Y}_{v}))+{b}_{1}, \\
  \hat{y} &= softmax({W}_{2}{P}_{tav}+{b}_{2}),
  \end{split}
\end{equation}    
where ${W}_{1}$, ${W}_{2}$, ${b}_{1}$ and ${b}_{2}$ are trainable parameters. We apply the standard cross-entropy loss function to train the model, for ${N}_{b}$ utterances in a batch, these are calculated as:
\begin{equation}
  \mathcal{L}=-\frac{1}{{N}_{b}}\sum^{{N}_{b}}_{i=0}{y_i} \cdot log{\hat{y}_{i}}.
\end{equation}

\section{Experiments}
\label{sec:Experiments}

\begin{table*}[t]
 \caption{
 We present the overall performance of Acc-7 and w-F1. We also present w-F1 scores for each category except for the \textit{Disgust} and \textit{Fear} categories, where the sample is smaller. Bold marks indicate best performance.
 }
  \centering
  \begin{tabular}{c ccccc cc}
  \toprule
  \multicolumn{1}{c}{\multirow{2}{*}{Methods}} & \multicolumn{7}{c}{MELD}
  \\ 
  \cline{2-8}
   & \textit{Anger} & \textit{Joy} & \textit{Neutral}  
   &  \textit{Sadness}
   &\textit{Surprise}  & \textbf{{Acc-7}} & \textbf{{w-F1}}    \\
  \midrule
    TFN \cite{DBLP:conf/emnlp/ZadehCPCM17}
    & 44.15 & 51.28 & 77.43 & 18.06 & 47.89 & 60.77 & 57.74 \\ 
    LMF \cite{DBLP:conf/acl/MorencyLZLSL18}
    & 46.64 & 54.20 & 76.97 & 21.15 & 47.06 & 61.15 & 58.30 \\
    DialogueCRN \cite{DBLP:conf/acl/HuWH20} 
    & 45.15 & 52.77 & 77.01 & 26.63 & 50.10 & 61.11 & 58.67  \\
    DialogueGCN \cite{DBLP:conf/emnlp/GhosalMPCG19} 
    & 40.83 & 51.20 & 75.97 & 19.60 & 46.05 & 	59.46   & 	58.10   \\
    MMGCN \cite{DBLP:conf/acl/HuLZJ21}
    & 46.09 & 53.02 & 76.33 & 26.74 & 48.15 & 60.42 & 58.65 \\
    MM-DFN \cite{DBLP:conf/icassp/HuHWJM22}
    & \textbf{47.82} & \textbf{54.78} & 77.76 & 22.93 & 50.69 & 62.49 & 59.46 \\
    \hline 
    \textbf{AIMDiT}    
    & 44.99 & 54.00 &  \textbf{80.24}  &  \textbf{27.75}  &  \textbf{56.97} & 	\textbf{64.83}   & 	\textbf{62.33}   \\ \bottomrule
    
  \end{tabular}    
   \label{tab:overall}
\end{table*}

\subsection{Datasets}
We conducted our experiments using the \textbf{MELD} dataset \cite{DBLP:conf/acl/PoriaHMNCM19}.
MELD contains 13708 utterances out of 1433 dialogues of the TV series Old Friends.
The dataset contains seven emotions, namely Anger, Disgust, Fear, Happy, Neutral, Sadness and Surprise. 
For a fair comparison, we conducted experiments using the MELD dataset pre-set with training, validation and test splits.

\begin{table}[t]
  \caption{
  The results under different combinations of modes are compared in terms of Acc-7 and w-F1 metrics.
  }
  \centering
      \resizebox{1.0\linewidth}{!}{$   
    \begin{tabular}{p{4cm}<{\centering} p{1.9cm}<{\centering} p{1.9cm}<{\centering}}
    \toprule
      Modality & Acc-7 & w-F1 \\
      \midrule
       Text           & 62.83          & 60.15           \\ 
       Audio            & 48.46          & 32.39           \\  
       Visual    & 48.12          & 31.26           \\  
       Text + Audio    & 63.12          & 60.38           \\  
       Text + Visual          & 62.37              & 60.71           \\  
       Audio + Visual         & 49.54          & 35.46           \\  
       Text + Audio + Visual         & \textbf{64.83}      & \textbf{62.33}           \\  
      \bottomrule
    \end{tabular}      
   $}

\label{tab:modality}
\end{table}

\begin{table}[t]
  \caption{
  Ablation Study of the MAN and MIN module. 
  }
  \centering
      \resizebox{1.0\linewidth}{!}{$   
    \begin{tabular}{p{4cm}<{\centering} p{1.9cm}<{\centering} p{1.9cm}<{\centering}}
    
      \toprule
      AIMDiT & Acc-7 & w-F1 \\
      \midrule
       -w/o MAN            & 61.72          & 59.54           \\  
       -w/o MIN            & 62.91          & 59.62           \\  
       \textbf{AIMDiT (Full)}    & \textbf{64.83}     & \textbf{62.33}      \\  
      \bottomrule
    \end{tabular}      
   $}
\label{tab:modules}
\end{table}

\subsection{Comparison Methods}

\textbf{TFN}\cite{DBLP:conf/emnlp/ZadehCPCM17} utilizes an early fusion approach for modal fusion. \textbf{LMF}\cite{DBLP:conf/acl/MorencyLZLSL18} low ranks the tensor for fusion. \textbf{DialogueGCN}\cite{DBLP:conf/emnlp/GhosalMPCG19} utilizes the dependency of graph structural context to model conversational data. \textbf{DialogueCRN}\cite{DBLP:conf/acl/HuWH20} designs multi-turn reasoning modules to understand conversational context. \textbf{MMGCN}\cite{DBLP:conf/acl/HuLZJ21} utilizes the structural features of graphs to capture intra- and inter-modal features.  \textbf{MM-DFN}\cite{DBLP:conf/icassp/HuHWJM22} designs graph-based dynamic fusion modules for multimodal context fusion in conversations, and reduces redundant information by capturing context dynamics. Note that \cite{DBLP:conf/emnlp/GhosalMPCG19} and \cite{DBLP:conf/acl/HuWH20} are designed for single-modality, thus introducing early fusion as a multimodal fusion method.

\textit{\textbf{Implementation Details.}}
For text, we used RoBERTa \cite{DBLP:conf/corr/LiuOGDJCLLZS19} to pre-train the model. For audio, we extracted Mel-spectrogram audio features via librosa. For visual, we used effecientNet \cite{DBLP:conf/icml/TanL19} pre-trained on the VGGface and AFEW datasets \cite{DBLP:conf/ivc/KossaifiTTP17, DBLP:conf/multi/DhallGLG12} to extract visual features.
For the dataset MELD, the batch size is 96, the initial learning rate is 1e-4.

\begin{table}[t]
  \caption{
  Fixed MIN and different performance of AIMDiT in different MAN layers. 
  }
  \centering
      \resizebox{1.0\linewidth}{!}{$   
    \begin{tabular}{p{2cm}<{\centering} p{2cm}<{\centering} p{1.9cm}<{\centering} p{1.9cm}<{\centering}}
      \toprule
      MAN & MIN & Acc-7 & w-F1 \\
      \midrule
       1            & 4          & \textbf{64.86}      & 61.64     \\  
       2            & 4          & 64.83       &\textbf{62.33}    \\  
       3    & 4          & 63.14      & 59.79     \\  
      \bottomrule
    \end{tabular}      
   $}

\label{tab:layers}
\end{table}

\subsection{Overall Results}
We compare our model with various state-of-the-art methods and the overall results are shown in Table~\ref{tab:overall}. It can be seen that AIMDiT outperforms the previous methods in terms of accuracy and F1 score on the MELD dataset. Compared to the previous SOTA, AIMDiT improves Acc-7 and w-F1 of MELD by 2.35 \% and 2.87 \% respectively. As can be seen from the results, our model outperforms a number of sophisticated fusion mechanisms such as TFN and LFN, as well as a range of graph-based fusion models, which illustrates the importance of multimodal fusion.

\subsection{Ablation Study}

\noindent
\textbf{Comparison under Different Modality Settings}
Table~\ref{tab:modality} shows a comparison of performance under different modal combinations. It can be seen that the bimodal and trimodal models are generally better than the unimodal model. In the bimodal model, the combination of text and audio (T+A) shows the best performance, indicating that text and audio features are more complementary. In contrast, audio and visual (A+V) have the worst effect, reflecting a high degree of distribution difference and redundancy in audio and visual features, which is also worth improving in the future.

\noindent
\textbf{Impact of modules.}
The MAN module has the ability to extract inter-modal and intra-modal features, and MIN has the ability to fuse the two features, in order to explore the importance of these two modules, we conducted ablation experiments on both modules and the results are shown in the Table~\ref{tab:modules}. 
It can be seen that after removing MAN and MIN, compared with the complete model, there is a significant decrease in the performance of both, especially after removing MAN there is a decrease of more than 3 \%, which reflects the ability of MAN to extract features.

\noindent
\textbf{Impact of MAN layers}
In order to study the effect of different layers of MAN, we keep the number of MIN layers as 4 and change only the number of MAN layers. The experimental results are shown in Table~\ref{tab:layers}. We find that although the Acc-7 index is highest when the MAN has only 1 layer, the boost is small. Interestingly, the overall results are best when the MAN has 2 layers. However, increasing to 3 layers significantly reduces the effect. This indicates that redundant information between modes accumulates after a certain point in the number of layers.

\section{Conclusion}
\label{sec:typestyle}

In this paper, we proposed a novel multimodal fusion framework (AIMDiT) composed of Modality Augmentation Network (MAN) and Modality Interaction Network (MIN) for ERC tasks. Specifically, we designed a inter- and intra- modal features learning network MAN through dimension transformation and Inception convolution, and proposed MIN effectively integrates inter- and intra-modal features for interaction. Extensive experiments on the benchmark dataset MELD demonstrate the effectiveness and superiority of AIMDiT.




\bibliographystyle{IEEEbib}
\bibliography{strings,refs}

\end{document}